\documentclass[english,PRL,two column,superscriptaddress,amsmath]{revtex4-1}
\usepackage[T1]{fontenc}
\usepackage[latin9]{inputenc}
\usepackage{babel}
\usepackage{amstext}
\usepackage{graphicx}
\usepackage[unicode=true,pdfusetitle,
 bookmarks=true,bookmarksnumbered=false,bookmarksopen=false,
 breaklinks=false,pdfborder={0 0 1},backref=false,colorlinks=false]
 {hyperref}
\hypersetup{
 colorlinks,linkcolor=red,citecolor=blue}
\usepackage{breakurl}

\makeatletter

\@ifundefined{textcolor}{}
{%
 \definecolor{BLACK}{gray}{0}
 \definecolor{WHITE}{gray}{1}
 \definecolor{RED}{rgb}{1,0,0}
 \definecolor{GREEN}{rgb}{0,1,0}
 \definecolor{BLUE}{rgb}{0,0,1}
 \definecolor{CYAN}{cmyk}{1,0,0,0}
 \definecolor{MAGENTA}{cmyk}{0,1,0,0}
 \definecolor{YELLOW}{cmyk}{0,0,1,0}
}

\makeatother

\begin{document}

\preprint{This line only printed with preprint option }

\title{Chiral Symmetry Breaking in Micro-Ring Optical Cavity By Engineered
Dissipation}

\author{Fang-Jie Shu}

\address{School of Physics and Electrical Information, Shangqiu Normal University,
Shangqiu, Henan 476000, China}

\address{Key Laboratory of Quantum Information, University of Science and
Technology of China, CAS, Hefei, Anhui 230026, China}

\address{Electrical and Systems Engineering Department, Washington University,
St. Louis. Missouri 63130, USA}

\author{Chang-Ling Zou}

\thanks{Corresponding author: clzou321@ustc.edu.cn}

\address{Key Laboratory of Quantum Information, University of Science and
Technology of China, CAS, Hefei, Anhui 230026, China}

\address{Synergetic Innovation Center of Quantum Information \& Quantum Physics,
University of Science and Technology of China, Hefei, Anhui 230026,
China}

\address{Department of Applied Physics, Yale University, New Haven, Connecticut
06511, USA}

\author{Xu-bo Zou}

\address{Key Laboratory of Quantum Information, University of Science and
Technology of China, CAS, Hefei, Anhui 230026, China}

\address{Synergetic Innovation Center of Quantum Information \& Quantum Physics,
University of Science and Technology of China, Hefei, Anhui 230026,
China}

\author{Lan Yang}

\address{Electrical and Systems Engineering Department, Washington University,
St. Louis. Missouri 63130, USA}
\begin{abstract}
We propose a method to break the chiral symmetry of light in traveling
wave resonators by coupling the optical modes to a lossy channel.
Through the engineered dissipation, an indirect dissipative coupling
between two oppositely propagating modes can be realized. Combining
with reactive coupling, it can break the chiral symmetry of the resonator,
allowing light propagating only in one direction. The chiral symmetry
breaking is numerically verified by the simulation of an electromagnetic
field in a micro-ring cavity, with proper refractive index distributions.
This work provokes us to emphasize the dissipation engineering in
photonics, and the generalized idea can also be applied to other systems.
\end{abstract}
\maketitle

\section{introduction}

Usually, traveling wave resonators, such as whispering-gallery (WG)
microcavities \citep{Chiasera2010,Foreman2015,Yang2015}, hold the
mirror reflection geometry symmetry and support clockwise (CW) and
counter clockwise (CCW) traveling wave modes simultaneously. Because
of time-reverse symmetry of light, CW and CCW modes degenerate in
non-gyromagnetic materials. When there is external perturbation to
the cavity boundary, the CW and CCW modes will couple to each other
and form a pair of standing wave modes, which are odd and even symmetric
superpositions of CW and CCW waves with equal weights \citep{Kippenberg2002,Wu2009}.
In consequence, transmission of a passive cavity or the emission of
an active cavity are symmetric in both directions. Although this feature
has no harm for most applications like narrow-band filters, optical
resonators with chiral symmetry breaking are of great interest in
both physics and many applications including unidirectional lasers
\citep{redding2012local}.

Several approaches have been proposed to break the chiral symmetry
of a microcavity system. One category of approaches addresses the
issue by deforming a micro-cavity to break the chiral symmetry in
terms of geometry. People can create deformed microcavities with a
chiral boundary shape, such as a spiral micro-cavity \citep{chern2003unidirectional,song2014combination}
and an asymmetric limacon micro-cavity \citep{wiersig2008combining},
to realize unidirectional emission of light. A circular cavity can
also be tuned to a chiral one with multiple scatterers on the rim
\citep{zhu2010controlled,wiersig2011structure,kim2014partially,wiersig2014enhancing,wiersig2016sensors},
which will benefit microcavity-based nanoparticle sensing. In another
category, a rotation \citep{terrel2009performance,ge2014rotation}
or a time dependent modulation \citep{sounas2013giant,dong2015brillouin,kim2015non}
along angular direction of a microcavity introduces an effective angular
momentum, breaking the time reversal symmetry of light in the WG microcavity.
Furthermore, both the geometry and time symmetry are broken by rotating
a limacon cavity \citep{sarma2015rotating}, which provides a sensitive
on-chip rotation detector by monitoring far-field emission patterns.

In this paper, we propose an alternative approach to break the chiral
symmetry in a micro-ring cavity, based on the engineered dissipation
\citep{peng2014loss,golshani2014impact,longhi2015non,metelmann2015nonreciprocal}
of optical modes. By balancing the direct reactive coupling and indirect
dissipative coupling with proper phase difference, the cavity eigenmodes
become pure CW or CCW, which are the chiral modes in a micro-ring.
To verify the theory, we introduce dielectric perturbations to a normal
micro-ring, that is, adding pure real or imaginary dielectric constant
onto original one.

The numerical simulation based on the Maxwell equations confirms the
prediction of our analytical model. The chiral eigenmodes and an asymmetric
transmission spectra in forward and backward directions are shown
in the micro-ring system. We believe this simple and dissipative-related
asymmetry micro-ring can be used as a laser resonator with unidirectional
emission and as a key element of a highly sensitive sensor \citep{baaske2014single,shen2016detection}.

\begin{figure}
\includegraphics[width=8cm]{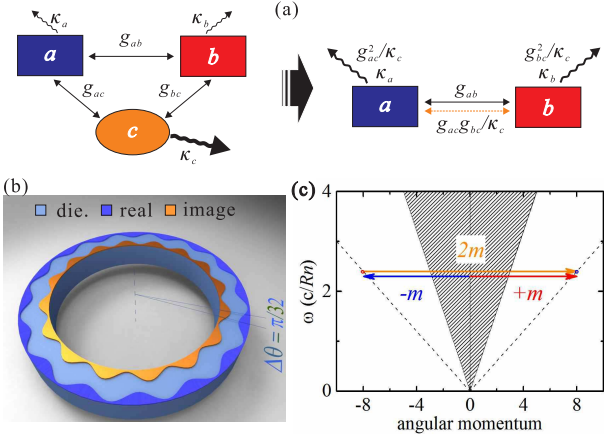}

\protect\caption{(Color online) (a) A schematic illustration of a dissipative coupling
by an engineered dissipation. The modes $a$ and $b$ couple to a
common lossy channel $c$ (left penal). It can be reduced to a effective
model in which $a$ and $b$ dissipatively coupled with each other
(right penal). (b) A micro-ring with dielectric and absorptive materials
depositing or doping on each rims works as a pure chiral resonator.
(c) A phase matching diagram for modes with mode index $m=\pm8$.}
\label{sketch}
\end{figure}

\section{Principle}

Before a detailed study on the WG modes, we briefly discuss a general
model of reactive and dissipative interactions between bosonic modes.
Suppose there are two ordinary mutual coupled modes $a$ and $b$
coupling with a lossy mode $c$, then the coupled mode equations described
the three mode system are (see Fig. \ref{sketch}(a) left penal)
\begin{equation}
\frac{d}{d\zeta}\left(\begin{array}{c}
a\\
b\\
c
\end{array}\right)=\left(\begin{array}{ccc}
\chi_{a} & ig_{ab} & ig_{ac}\\
ig_{ab} & \chi_{b} & ig_{bc}\\
ig_{ac} & ig_{bc} & \chi_{c}
\end{array}\right)\left(\begin{array}{c}
a\\
b\\
c
\end{array}\right).
\end{equation}
Here, the system evolves along the general axis $\zeta$, which can
be time, space or other generalized coordinates. In the parameter
$\chi_{x}=-i\omega_{x}-\kappa_{x}$, $\omega_{x}$ and $\kappa_{x}$
are intrinsic oscillation frequencies and rates of dissipation to
independent channels, respectively. They vary with coordinate $\zeta$
for mode $x\in\{a,b,c\}$. The $g_{ab}$, $g_{ac}$ and $g_{bc}$
are the common reactive couplings between modes a, b, and c, which
implies the coupling inducing a coherent energy transfer between the
modes. If $\kappa_{c}\gg max(\kappa_{a},\,\kappa_{b},\, g_{ab},\, g_{ac},\, g_{bc})$
and $\kappa_{c}\gg max(\omega_{c},\,|\omega_{c}-\omega_{a}|,\,|\omega_{c}-\omega_{b}|)$
are satisfied, i.e. relaxation rate of the mode $c$ is ultra-fast
and the mode $c$ is approximately at the static state $\frac{d}{d\zeta}c=0$.
Therefore, $c\approx\frac{ig_{ca}}{\kappa_{c}}a+\frac{ig_{cb}}{\kappa_{c}}b$
and we arrive at the effective coupling modes equations (Fig. \ref{sketch}(a)
right penal)
\begin{equation}
\frac{d}{d\zeta}\left(\begin{array}{c}
a\\
b
\end{array}\right)=\left(\begin{array}{cc}
\chi_{a}-\frac{g_{ac}^{2}}{\kappa_{c}} & ig_{ab}-\frac{g_{ac}g_{bc}}{\kappa_{c}}\\
ig_{ab}-\frac{g_{ac}g_{bc}}{\kappa_{c}} & \chi_{b}-\frac{g_{bc}^{2}}{\kappa_{c}}
\end{array}\right)\left(\begin{array}{c}
a\\
b
\end{array}\right).\label{eq:2}
\end{equation}
As expected, the introduced lossy channel to modes $a$ and $b$ induces
additional dissipations $\frac{g_{ac}^{2}}{\kappa_{c}}$ and $\frac{g_{bc}^{2}}{\kappa_{c}}$,
respectively. However, it is not intuitive that there is an additional
term $\frac{g_{ac}g_{cb}}{\kappa_{c}}$ for the coupling between the
two modes, due to the coherent interference in the shared lossy channel.
The additional coupling term is a real number, which indicates that
it is a dissipative coupling instead of a reactive coupling.

If the the coupling to bath ($g_{ac,bc}$) and the internal coupling
($g_{ab}$) are modulated along $\zeta$, the effective coupling between
$a$ and $b$ can be rewrote as complex function of $\zeta$
\begin{equation}
g_{eff}(\zeta)=|g_{eff}(\zeta)|e^{i\varphi(\zeta)}.
\end{equation}
The nontrivial phase accumulation $\varphi(\zeta)$ will lead to interesting
effects, which are absent in a system of pure reactive or dissipative
coupling. In this paper, we focus on a specially engineered dissipation
in which $\varphi(\zeta)=\Omega\zeta$ is a linear equation of $\zeta$.
As a consequence, the effective coupling $|g_{eff}(\zeta)|e^{i\Omega\zeta}$
includes a generalized momentum $\Omega$, which breaks $\zeta$-symmetry
of a system.

\section{Chiral symmetry breaking}

Now, we apply the principle to the WG micro-ring as depicted in Fig.$\,$\ref{sketch}(b),
where $\zeta$ denotes an angle $\theta$. Mode amplitudes in WG micro-ring
evolve as
\begin{equation}
\frac{d}{d\theta}a_{m}=(-im-\frac{m}{2Q})a_{m}=\chi_{m}^{(\theta)}a_{m},
\end{equation}
where $m$ is an integer corresponding to the angular momentum (mode
index) of the mode, and $Q$ is the quality factor of the mode. The
modes with positive and negative $m$ have a degenerate mode frequency
but opposite propagation directions, which are called clockwise (CW)
and counter-clockwise (CCW) modes. To obtain an effective coupling
between CW and CCW modes with the non-zero momentum $\Omega$, extra
materials distributed sinusoidally on the inner and the outer rims
are constructed, noting that there is a shifted angle $\Delta\theta$
between these two perturbations (Fig.$\,$\ref{sketch}(b)). The reactive
coupling between traveling wave modes is controlled by external refractive
index perturbation (dark blue part on the micro-ring in Fig.$\,$\ref{sketch}(b)),
such as a dielectric pattern on the surface of the micro-ring. The
dissipative coupling can be introduced by putting lossy nanoparticle
or doping absorptive atom on the cavity (orange part on the micro-ring
in Fig.$\,$\ref{sketch}(b)), which induces losses to CW and CCW
modes simultaneously.

\begin{figure*}
\includegraphics[width=14cm]{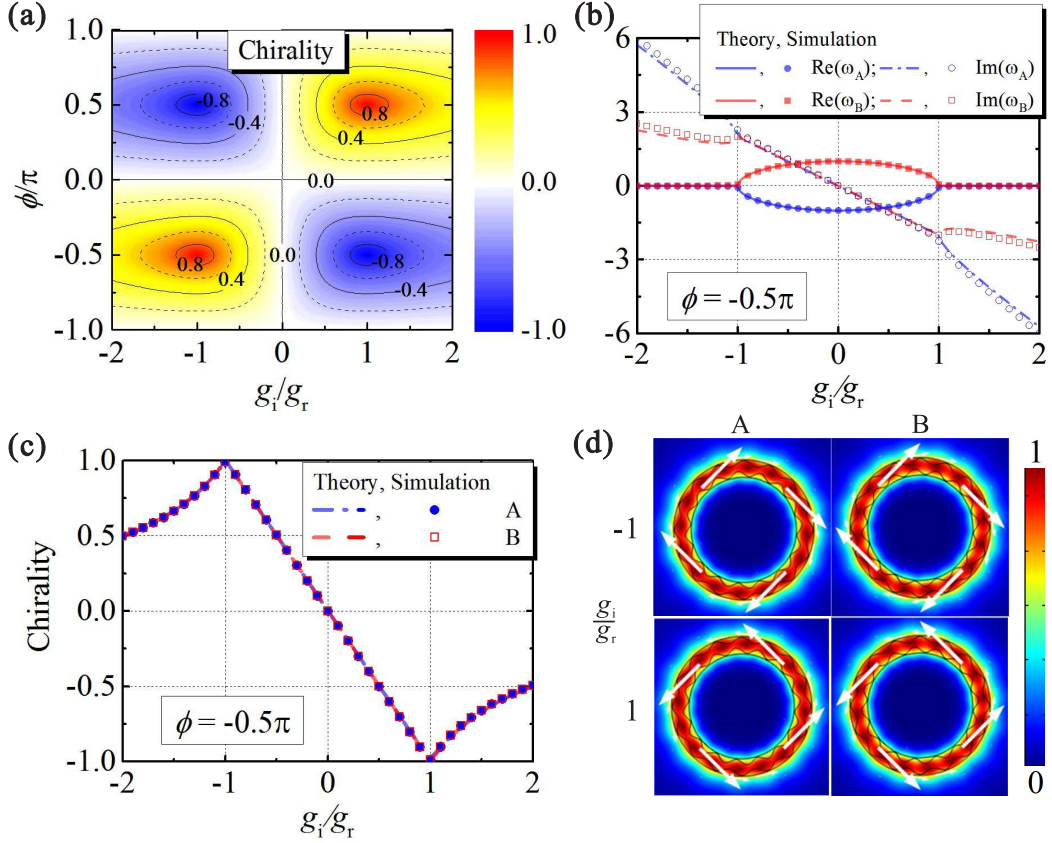}

\protect\caption{(Color online) (a) A chirality $K$ as a function of $g_{i}/g_{r}$
and $\phi$. The frequencies (b) and chirality (c) of eigenmodes in
the chiral microresonator with $\phi=-\frac{\pi}{2}$. The lines are
analytical results, and dots are from numerical simulations. (d) The
mode field presents with modulus of electric field intensity distributions
of $K=+1$ (first row) and $-1$ (last row), where white arrows indicate
the Poynting vectors.}
\label{dw}
\end{figure*}

We can formalize the effective dielectric perturbation as
\begin{eqnarray}
\delta\varepsilon(\theta) & = & \delta\varepsilon_{r}[\cos\Omega(\theta+\Delta\theta)+1]+i\delta\varepsilon_{i}(\cos\Omega\theta+1)\nonumber \\
 & = & \delta\varepsilon_{+}e^{i\Omega\theta}+\delta\varepsilon_{-}e^{-i\Omega\theta}+\delta\varepsilon_{0},
\end{eqnarray}
where $\delta\varepsilon_{+}=\frac{\delta\epsilon_{r}e^{i\phi}+i\delta\epsilon_{i}}{2}$,
$\delta\varepsilon_{-}=\frac{\delta\varepsilon_{r}e^{-i\phi}+i\delta\varepsilon_{i}}{2}$,
$\delta\varepsilon_{0}=\delta\varepsilon_{r}+i\delta\varepsilon_{i}$
and $\phi=\Omega\Delta\theta$. Due to the phase matching condition,
the term $e^{i\Omega\theta}$ scatters the light in the mode $m$
to the mode $m+\Omega$. For example, in Fig.$\,$\ref{sketch} (b)
we have a created momentum $\Omega=16$ by engineering the dissipative
coupling. This coupling couples $m=-8$ mode to $m=8$ mode ((Fig.$\,$\ref{sketch}(c))).
It is worth noting that besides the phase matching condition, energy
conservation is another criterion for effective coupling. As a result,
mode coupling occurs between degeneracy modes with a same $|m|$.
For example, the phase matching condition between $m=-7$ and $m=9$
is satisfied, but their mode frequencies are different by two free
spectra ranges (FSR), which is many orders of magnitude larger than
the intrinsic linewidth of the high-$Q$ WG modes. So they cannot
couple with each other in this way.

In the following, we take the $m=\pm8$ modes in the micro-ring (Fig.$\,$\ref{sketch}(b))
as an example, and denote the CW and CCW modes with $a_{+}$ and $a_{-}$
for simplicity. The evolution of two modes, a specific form of the
Eq. (\ref{eq:2}), can be written as
\begin{equation}
\frac{d}{d\theta}\left(\begin{array}{c}
a_{+}\\
a_{-}
\end{array}\right)=\left(\begin{array}{cc}
\chi_{m}^{(\theta)}-\varrho\delta\varepsilon_{0} & \varrho\delta\varepsilon_{+})\\
\varrho\delta\varepsilon_{-} & \chi_{-m}^{(\theta)}-\varrho\delta\varepsilon_{0}
\end{array}\right)\left(\begin{array}{c}
a_{+}\\
a_{-}
\end{array}\right).
\end{equation}
Here, perturbation effect is proportional to the $\delta\varepsilon(\theta)$
with a coefficient $\varrho$. Since the light traveling in the micro-ring
is with periodic boundary condition, we translate the coordinate $\theta$
to $t=\tau_{r}\theta/2\pi$ for convenience, where the $\tau_{r}$
is the round-trip time of the cavity. Then, in a rotating frame of
cavity frequency, the temporal evolution becomes
\begin{equation}
\frac{d}{dt}\left(\begin{array}{c}
a_{+}\\
a_{-}
\end{array}\right)=\left(\begin{array}{cc}
-\kappa-2g_{i} & ig_{r}e^{i\phi}-g_{i}\\
ig_{r}e^{-i\phi}-g_{i} & -\kappa-2g_{i}
\end{array}\right)\left(\begin{array}{c}
a_{+}\\
a_{-}
\end{array}\right),
\end{equation}
where $g_{r}$ and $g_{i}$ are the reactive and dissipative coupling
strength, respectively.

The normal modes of the coupling system can be solved as
\begin{eqnarray}
A,B & = & \frac{\pm pa_{+}+a_{-}}{\sqrt{1+|p|^{2}}},\label{eq:eigenvector}
\end{eqnarray}
where
\begin{equation}
p=\frac{ie^{i\frac{\phi}{2}}\sqrt{(ig_{r}-g_{i}e^{i\phi})(g_{i}-ig_{r}e^{i\phi})}}{ig_{r}-g_{i}e^{i\phi}}.
\end{equation}
The eigenfrequencies are
\begin{eqnarray}
\omega_{A,B} & =-i2g_{i}\mp & e^{-i\frac{\phi}{2}}\sqrt{(ig_{r}-g_{i}e^{i\phi})(g_{i}-ig_{r}e^{i\phi})}.\label{eq:eigenfrequency}
\end{eqnarray}

From the Eq.$\,$(\ref{eq:eigenvector}), the ingredients of CW and
CCW circulating light in the new normal modes are generally not balanced.
For both normal modes, the portion of CCW light is $\frac{1}{1+|p|^{2}}$.
Here, we define the chirality of the system as
\begin{equation}
K=\frac{I_{CW}-I_{CCW}}{I_{CW}+I_{CCW}}=\frac{|p|^{2}-1}{|p|^{2}+1},\label{eq:K}
\end{equation}
where $I_{CW}$($I_{CCW}$) is energy of CW(CCW) ingredient.

In Fig. \ref{dw}(a), the chirality $K$ of the modes are theoretically
calculated with respect to relative coupling strengths $\frac{g_{i}}{g_{r}}$
and phase differences $\phi$. It is clearly shown that $K=0$ if
$g_{i}$ or $\phi$ is zero, and high chirality is obtained near the
points $(\pm1,\pm\pi/2)$ . Specially, if we choose $\phi=-\frac{\pi}{2}$,
analytical results will be simplified as $p=i\sqrt{\frac{g_{i}}{g_{r}}-1}/\sqrt{\frac{g_{i}}{g_{r}}+1}$.
As shown in Fig.$\,$\ref{dw}(b), the system behaviors have two regions
separated by points $|\frac{g_{i}}{g_{r}}|=1$. In region $\frac{g_{i}}{g_{r}}\in[-1,1]$,
the frequency splits, i.e. $\mathrm{Re}(\omega_{A}-\omega_{B})\neq0$,
while the imaginary part are the same $-i2g_{i}$, thus the couplings
only change the frequencies of the uncoupled degenerated CW and CCW
modes. For parameter $\frac{g_{i}}{g_{r}}$ outside the interval,
$\omega_{A,B}$ are pure imaginary numbers, which removes the degeneracy
by adding different losses. Remarkably, for $\frac{g_{i}}{g_{r}}=1$
($\frac{g_{i}}{g_{r}}=-1$), the system only supports pure CCW (CW)
mode as $K=-1$ ($+1$) (Fig. \ref{dw}(c)).

\section{Numerical Results}

To check the validity of the idea about breaking chiral symmetry in
traveling wave optical resonators by engineering the dissipation,
we solve out eigenmodes and reflection spectra in those cavities numerically.
Such numerical verification is necessary, because the principle and
model in previous section are idealized. In realistic systems, there
should be multiple modes, and the coupled mode equations are just
 results of perturbation theory about the system, these approximations
may lead to unpredicted effect to the chiral symmetry breaking.

In the following, the model proposed in Fig.$\,$\ref{sketch}(b)
is simulated by finite element method (COMSOL multiphysics). In the
simulation, only the geometry and the dielectric constant of the ring
distribution are set in the software. The eigenmodes and propagations
of light are obtained by using the numerical method to solve the Maxwell
equations, without any assumptions. Therefore, the numerical results
in the following part will provide an independent and experiment-feasible
verification to our theory.

\subsection{Eigenfrequencies and Chirality}

The geometry of model in COMSOL is the same as that in Fig. \ref{sketch}(b),
but simplified in two-dimensional model. In this model, the radii
of inner and outer circle are $0.6\,\mathrm{\mu m}$ and $0.8\,\mathrm{\mu m}$,
respectively, so the width of the ring is $0.2\,\mathrm{\mu m}$.
In addition, amplitudes of both sinusoidal curves are $0.03\,\mathrm{\mu m}$,
and shift $0.01\,\mathrm{\mu m}$ away from rims to avoid sharp structure
which arises singularity in the simulation. In the material side,
the dielectric constant of the main body of the ring is $\epsilon=10$.
The inner and outer parts sinusoidal curves inclosed dielectric perturbation
($\delta\epsilon$) of imaginary ($\delta\epsilon$$_{i}$) and real
($\delta\epsilon$$_{r}$) part, respectively. The $\delta\epsilon$$_{i}$
and the $\delta\epsilon$$_{r}$ are three orders of magnitude less
than $\epsilon$ to insure the validation of the perturbation approximation.
As mentioned in the above analysis, the coupling strength $g_{i}$
($g_{r}$) should be proportional to the perturbation $\delta\epsilon$$_{i}$
($\delta\epsilon$$_{r}$) \citep{saleh1991fundamentals}. Due to
the asymmetric field distribution in the inner and outer rim of the
micro-ring resonator, the balanced reactive and dissipative couplings
are achieved with $\delta\epsilon_{i}=0.01$ and $\delta\epsilon_{r}=0.005175$.
It should be noticed that the sign of $\delta\epsilon_{i}$ determines
whether the materials loss ($+$) or gain ($-$). The sign of $\delta\epsilon_{i}$
changes easily either in theory or in simulation, but in practical
experiments loss is preferred for convenience.

We performed the simulation with fixed $\phi=-\frac{\pi}{2}$ and
$\delta\epsilon_{r}=0.005175$ but varies $\delta\epsilon_{i}$ from
$0.02$ to $-0.02$ corresponding to that $\frac{g_{i}}{g_{r}}$ varies
from $-2$ to $2$. To compare with the theory, the numerical results
of $\omega$ are drawn in coordinate system with original point $(\omega_{A}-\omega_{B})/2$
and frequency unit $|\textrm{real}(\omega_{A}-\omega_{B})|/2$, where
$\omega_{A}$ and $\omega_{B}$ are evaluated at $g_{i}=0$. It is
shown in Fig. \ref{dw}(b) that numerical results (symbols) are well
consistent with analytic one (line). In addition, the chirality can
calculated numerically by
\begin{equation}
K=\pm2\frac{E_{max}E_{min}}{E_{max}^{2}+E_{min}^{2}},\label{eq:K-1}
\end{equation}
where $E_{min}$ and $E_{max}$ are minimum and maximum of electric
field modules along the path $r=0.72\,\mathrm{\mu m},$ (Fig. \ref{dw}(c)).
The prior circulate direction or sign of chirality is known from Poynting
vectors (white arrows in Fig. \ref{dw}(d)). The mode distributions
at point $\frac{g_{i}}{g_{r}}=1$ (Fig. \ref{dw}(d) lower panel)
show remarkable CCW traveling mode with $K\approx-1$. The reverse
CW mode can also be achieved at $\frac{g_{i}}{g_{r}}=-1$ (Fig. \ref{dw}(d)
upper panel) or at $\phi=\frac{\pi}{2}$ (Fig. \ref{dw}(a)).

\subsection{Reflection and Transmission Spectrum}

\begin{figure}
\includegraphics[width=8cm]{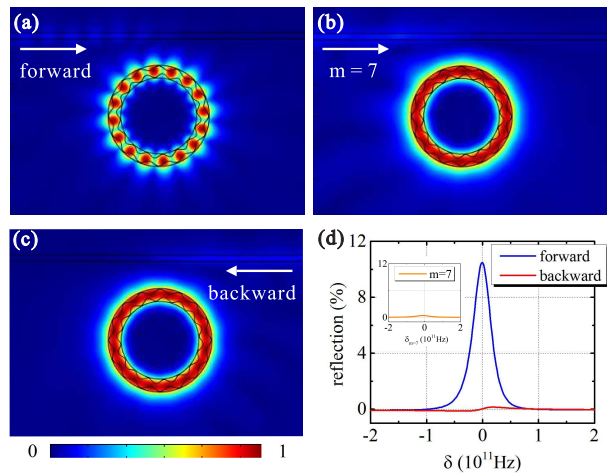}

\protect\caption{(Color online) (a) Forward coupling in critical coupling region with
$\delta=0$. A back scattered wave forms a standing wave in the cavity.
(b) Forward coupling of $m=7$ mode. (c) Backward coupling.The false
color represents modulus of electric field intensity. (d) Spectrum
of reflections $R$ in aforementioned situations.}

\label{trans}
\end{figure}

The chirality can also be checked by sending probe laser to the cavity
and measuring the reflectivity and transitivity of the cavity. As
shown in Fig.$\,$\ref{trans}, by putting a straight dielectric waveguide
in the vicinity of the micro-ring, light from either left or right
port in waveguide can couple with the modes in micro-ring. Suppose
that there are input signals $E_{in}^{+}=a_{in}^{+}e^{-i\omega_{in}t}$
and $E_{in}^{-}=a_{in}^{-}e^{-i\omega_{in}t}$ couple to CW and CCW
modes in the cavity, respectively. The coupling mode equations become
\begin{eqnarray}
\frac{da_{+}}{dt} & = & (-i\omega_{0}-\kappa-2g_{i}-\kappa_{in})a_{+}+(ig_{r}e^{i\phi}-g_{i})a_{-}\nonumber \\
 &  & +\sqrt{2\kappa_{in}}a_{in}^{+}e^{-i\omega_{in}t},\label{eq:CM1-1}\\
\frac{da_{-}}{dt} & = & (-i\omega_{0}-\kappa-2g_{i}-\kappa_{in})a_{-}+(ig_{r}e^{-i\phi}-g_{i})a_{+}\nonumber \\
 &  & +\sqrt{2\kappa_{in}}a_{in}^{-}e^{-i\omega_{in}t}.\label{eq:CM2-1}
\end{eqnarray}
We focus on the pure CCW microcavity with $\frac{g_{i}}{g_{r}}=1$
and $\phi=-\frac{\pi}{2}$. With the input-output relations $a_{out}^{+}=a_{in}^{+}-\sqrt{2\kappa_{in}}a_{+}$,
$a_{out}^{-}=a_{in}^{-}-\sqrt{2\kappa_{in}}a_{-}$, the forward steady-state
transmissions (coincident with CW, $a_{in}^{+}\neq0$) and the backwards
ones at critical coupling $\kappa_{in}=\kappa+2g_{i}=\gamma$ and
detuning $\delta=\omega_{in}-\omega_{0}$ are
\begin{eqnarray}
T_{+} & = & |\frac{a_{out}^{+}}{a_{in}^{+}}|^{2}=\frac{\delta^{2}}{\delta^{2}+4\gamma^{2}},\label{eq:Tb}\\
T_{-} & = & |\frac{a_{out}^{-}}{a_{in}^{-}}|^{2}=|\frac{i\delta}{i\delta-2\gamma}+\frac{a_{in}^{+}}{a_{in}^{-}}\frac{2g_{r}2\gamma}{(i\delta-2\gamma)^{2}}|^{2}.\label{eq:Tf}
\end{eqnarray}
They are typical Lorentz line shapes as same as those in a symmetric
microcavity when only one input port has an input signal. The reflections
are
\begin{eqnarray}
R_{+} & = & |\frac{a_{out}^{-}}{a_{in}^{+}}|^{2}=|\frac{a_{in}^{-}}{a_{in}^{+}}\frac{i\delta}{i\delta-2\gamma}+\frac{2g_{r}2\gamma}{(i\delta-2\gamma)^{2}}|^{2}\label{eq:Tf-1}\\
R_{-} & = & |\frac{a_{out}^{+}}{a_{^{-}in}}|^{2}=|\frac{a_{in}^{+}}{a_{in}^{-}}\frac{i\delta}{i\delta-2\gamma}|^{2}\label{eq:Tb-1}
\end{eqnarray}
 A notable result is the reflectivity of the backward direction with
$\delta=0$

\begin{equation}
R_{+}=\left(\frac{g_{r}}{\gamma}\right)^{2}=\left(\frac{g_{r}}{\kappa+2g_{r}}\right)^{2},\label{eq:Rb}
\end{equation}
which is not trivial because of the asymmetric backward coupling in
the microcavities. In the case of input signal from forward direction,
efficiency back coupling from CW to CCW mode constructs a partial
stationary mode in the cavity (Fig.$\,$\ref{trans}(a)). Then a relative
high reflection of $R_{+}\approx10\%$ in on-resonate situation (Fig.$\,$\ref{trans}
(d) blue line), which is coincident with Eq.$\,$(\ref{eq:Rb}). Therein
the loss of the mode (Fig.$\,$\ref{dw}(d)) is $\gamma/2\pi=108\,\mathrm{GHz}$.
When $g_{i}=0$, the separation of two modes in Fig.$\,$\ref{dw}(c)
is $2g_{r}/2\pi\approx68.5\,\mathrm{GHz}$. Substituting them into
Eq.$\,$(\ref{eq:Rb}), we get $R_{+}\approx10\%$ consistent with
our numerical results. As shown in Fig.$\,$\ref{trans}(d), bandwidth
of chirality is about $50\,\mathrm{GHz}$ in terms of reflection,
which is half of the resonance line width. Higher reflection can be
achieved by increasing coupling strength $g_{r}$, and the maximum
achievable $R_{+}=1/4$ for $g_{r}/\kappa=\infty$. In addition, we
also check the chiral symmetry breaking for $m=7$ mode. Due to the
mismatch of energy, reflection is negligible (Figs.$\,$\ref{trans}(b)
and (d) insert). In backward input situation, reflection is negligible
as well (Figs.$\,$\ref{trans}(c) and (d) red line).

\section{Discussion}

It is possible to demonstrate the proposed structure in experiment
by fabricating nanostrucutres on the top of a on-chip micro-ring resonator.
A similar technique had been demonstrated in Ref.$\,$\citep{feng2014single}.
It is also possible to demonstrate the idea in a microsphere or microtoroid
cavity, by putting absorptive metal nanoparticle on the surface to
induce dissipative coupling, and putting a dielectric nanoparticle
or nanotip to the surface to induce reactive coupling.

A limitation of the approach studied here is the induced dissipation
to the system. The symmetry breaking is realized with the penalty
that a part of energy losses into environment. For the cavities studied
here, the best theoretical asymmetry in spectrum is 1/4 according
to Eq. (\ref{eq:Rb}). In principle, such limitation can be mitigated
by replacing the dissipative coupling with a gaining one, or say,
by changing the sign of coupling. In this case, the symmetry breaking
is obtained without additional loss.

The results presented in Fig.$\,$\ref{dw} show similar behaviors
observed in the study of parity-time (PT) symmetry \citep{Chang2014,Peng2014}
and exceptional point (EP) \citep{peng2014loss} in optical cavities.
Actually, the perfect chirality is obtained at EP, which is common
in non-Hermitian systems. Although the evolution of $\Re\omega$ and
$\Im\omega$ show similar splitting phenomena with increasing or decreasing
$g_{i}/g_{r}$, the system is intrinsically different from the PT-symmetry
models \citep{Chang2014,Peng2014}. In the PT-symmetry models, there
is reactive coupling but the two modes have different loss/gain factors.
In our model, we have both reactive and dissipative coupling, while
the two mode have exactly the same frequency. It is also different
from recently proposed and demonstrated anti-PT symmetry \citep{ge2013antisymmetric,Peng},
where they have a pure dissipative coupling but have different mode
frequencies.

As described in Sec. II, because the principle for breaking the symmetry
is general, it can be applied to any coordinate $\zeta$. One straightforward
generalization is breaking the forward and backward symmetry for wave
propagating in two coupled straight waveguides. The wave is not limited
to be light. It can be any coherent wave source, such as microwave,
acoustic wave, spin wave and water wave. Another interesting generalization
is breaking time-reversal symmetry, i.e. $\zeta=t$. In this case,
if we apply temporal modulations on the dissipative and reactive coupling
between systems, the non-reciprocal wave propagation can be realized
\citep{metelmann2015nonreciprocal}. We should aware that this approach
for non-reciprocity does not conflict with the Lorentz reciprocal
theorem, because the temporal modulations are nonlinear effects which
lead to the non-reciprocity.

\section{conclusion}

In summary, an approach to break the chiral symmetry of light in a
traveling wave microresonator is proposed and verified by numerical
simulations. Though the discussion is focusing on the chiral symmetry
breaking in coordinate $\theta$, the same concept can be used to
break symmetry in coordinate $x$ if we straighten the ring into a
straight waveguide. This concept also can be used to break time symmetry
or others if we introduce two modulations on $t$ or other physical
quantities. In terms of $t$, a modulation on the refractive index
by $t$ in the entirety system together with another modulation in
phase and with the same magnitude modulation on loss (gain) can break
the time symmetry.
\begin{acknowledgments}
We thank Ming Li, Yan-Lei Zhang, Yong Yang, Jianming Wen, and Liang
Jiang for fruitful discussion. The work was supported by the Strategic
Priority Research Program (B) of the Chinese Academy of Sciences (Grant
No. XDB01030200), National Basic Research Program of China (Grant
Nos. 2011CB921200 and 2011CBA00200) and the National Natural Science
Foundation of China (Grant No. 61505195). F.J.S. is supported by Program
for Innovative Research Team (in Science and Technology) in University
of Henan Province (IRTSTHN No. 16IRTSTHN028), the State Scholarship
Fund from China Scholarship Council (No.201508410405), and the Young
Core Instructor Foundation from the Education Department of Henan
Province, China (2013GGJS-163).
\end{acknowledgments}

\end{document}